\documentclass[<preprint>]{elsarticle}
\usepackage{amsmath}
\usepackage{amssymb}
\usepackage{graphicx}
\usepackage{dcolumn}
\usepackage{bm}
\usepackage[symbol]{footmisc}

\def\be{\begin{equation}}
\def\ee{\end{equation}}
\def\bea{\begin{eqnarray}}
\def\eea{\end{eqnarray}}

\begin{document}
\title{Temperature inhomogeneity in non-equilibrium field theory for electrons in a nanowire: thermodynamic and transport properties}
\author{Yuan Gao\footnote[1]{Author to whom any correspondence should be addressed: yuan.gao@ufl.edu} and K.A. Muttalib}
\affiliation{Department of Physics, University of Florida, P.O. Box 118440, Gainesville, FL 32611-8440}

\begin{abstract}

A nanowire with its two ends fixed at two different temperatures by external baths is the simplest example of a fermionic system with a temperature inhomogeneity, and could be an easy platform to study thermodynamic and transport properties of a boundary-driven open quantum system. Starting with a temperature-dependent pseudo free energy derived from an exact reduced density matrix and assuming a small temperature gradient $\gamma$ across the wire,  we show within perturbation theory that electron dispersion relation and therefore the Fermi distribution becomes $\gamma$ and space coordinates dependent, leading to non-linear effects of the temperature inhomogeneity. In particular, we show that in the non-linear response regime, the Widemann-Franz Law for the ratio of thermal and electrical conductivities is generalized, and that the thermopower increases with increasing temperature gradient $\gamma$.

\end{abstract}
\maketitle

\newpage


\section{Introduction}
We consider a simple boundary-driven open quantum system \cite{Li12, LPS22, Bertini21, DSP12, WAL14, FM07}, a nanowire attached to two leads (baths) kept at two different fixed temperatures, with electrons flowing from the hot lead to the cold one through the nanowire. Looking at the central subsystem, the nanowire only, unlike zero dimensional systems (quantum dots, atomic junctions, molecular junctions), temperature inhomogeneities could be formed across it \cite{H09} after long times. Considering electron contributions only, a  class of high-efficiency thermoelectric nano-devices has been proposed that takes advantage of an interplay between the material and the thermodynamic parameters in a device, available in the non-linear regime \cite{hmn13, mh15, mm21}. However, in this non-linear regime the non-equilibrium electrical or thermal current in the nanowire is usually calculated assuming a fixed homogeneous temperature inside the conductor. This is because the mechanism accountable for the formation of temperature inhomogeneity is still unclear, and one has to work within linear response theory to avoid the explicit temperature dependent contributions \cite{luttinger64}. While there are works on time evolution under an initial temperature profile on the level of perturbation expansion \cite{langmann17} and exact solutions available for some expectation values in one-dimensional models that can be mapped on to conformal field theories \cite{moosavi1, moosavi2}, there is no systematic theoretical framework available to address the full time evolution in the presence of fixed thermal inhomogeneity in general quantum systems (beyond strictly one-dimension), either within the quantum master equation formalism \cite{LPS22, Bertini21} or using the Non-Equilibrium Green’s Function (NEGF) techniques \cite{ WAL14, rammer}. 

In a recent work \cite{gm04}, a toy model of non-interacting phonons in a nanowire with a linear temperature profile was considered as a 'proof of concept' of a general framework to study non-interacting phonons with thermal inhomogeneity in the non-linear response regime. Instead of guessing the form of the scattering process responsible for the temperature gradient, the exact reduced density matrix characterized by the temperature profile was written down and served as an indicator to find out the effective scattering term. The work suggested that within perturbation theory where the temperature gradient across the system  is assumed to be small, thermal inhomogeneity results in temperature dependent scattering, generating corrections to the bosonic Green's function and the local density; this eventually results in a change in the non-linear thermal current carried by phonons upon the attachment of leads on both ends of the sample. Here we adapt the framework for a  toy model of non-interacting \textit{fermions} to show that, unlike bosonic systems, thermal inhomogeneity makes the dispersion relation of fermions, and thus Fermi distribution function dependent on the temperature gradient as well as the space coordinates, which could be measured with diverse experimental techniques \cite{Freud22, Wyzu22}. In addition, when the sample is attached to two leads, electron transport is significantly expedited due to the local density gradient caused by the thermal gradient. Transport coefficients evaluated by using NEGF formalism show, in the non-linear response regime, a robust increase of the thermopower or Seebeck coefficient $S_{e}$, and the Lorenz number $\frac{\kappa}{T\sigma}$, where $\kappa$ is the thermal conductivity from electrons, and $\sigma$ is the electrical conductivity. This is also interesting because for fermi-liquids the ratio $\frac{\kappa}{T\sigma}$ is fixed, given by the Wiedemann–Franz law, and the Seebeck coefficient (or thermopower) is a given material property in the linear response regime \cite{Gour22}.  

Note that a nanowire is not one dimensional, and we consider a wide nanowire with many channels. However, our model of temperature inhomogeneity only along one direction allows us to consider an effectively one-dimensional system along the direction of the temperature profile. Although we assume a simple linear temperature profile across a simple conductor, this work can serve as a benchmark for any future endeavours on the broad effects of temperature inhomogeneity in more complex systems.

\section{Framework and set up}
\begin{figure}
\includegraphics[angle=0,width=0.45\textwidth]{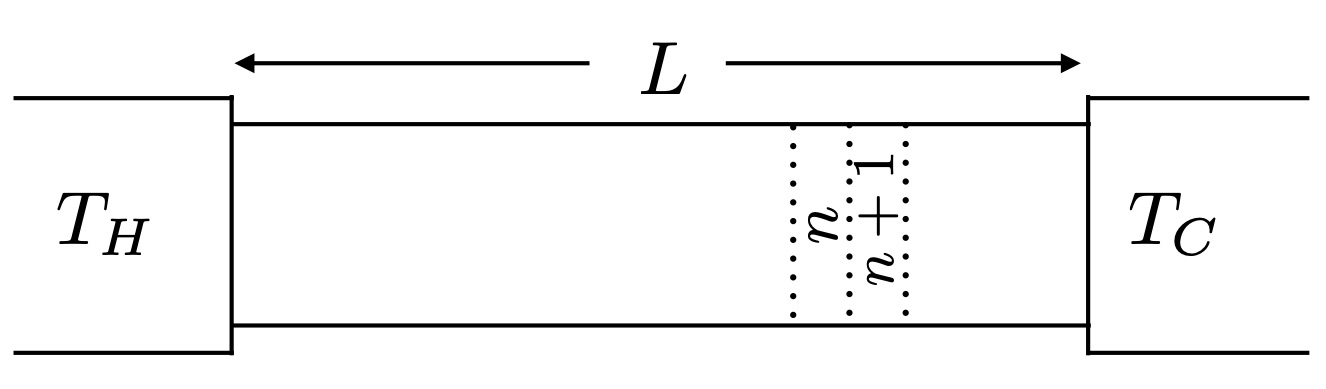}
\caption{Schematic setup for a temperature-inhomogeneous sample consisting of N subsystems coupled to two heat baths, taken from \cite{gm04}.}
\label{Fig-profile}
\end{figure}

The system of interest is a thermally inhomogeneous system with $N$ subsystems where temperature in each $n$th subsystem is different, as shown in Figure \ref{Fig-profile}. In the limited scope of this work, we focus on thermalizing systems where the temperature profiles are sustained by external bath and are fixed (quenched). This might include situations where, e.g.,  either the whole system is synthesized in a heat bath with temperature inhomogeneities and is thus in local equilibrium, or the system is brought out of equilibrium by means of coupling to a thermally inhomogeneous heat bath and eventually it  reaches a non-equilibrium steady state (NESS). In both cases, since the temperature profile is fixed,  it is possible to take into account the different temperatures at different positions after integrating out the bath degrees of freedom;  the thermally inhomogeneous system could be described as a combination of many subsystems, each with a fixed local temperature, by a reduced density matrix in conventional Gibbs form. In the absence of time dependent perturbations, both Hamiltonian and reduced density matrix are constant. However, the reduced density matrix of a thermally inhomogeneous system in general doesn't commute with a thermally homogeneous Hamiltonian. Conceptually this fails to describe systems whose expectation values of observables don't vary with time, which technically makes Wick’s theorem not directly applicable in the perturbation expansion of Green’s functions. In a previous work \cite{gm04} we proposed that the starting point of the problem should be the exact reduced density matrix $\rho(t_0)$, and the evolution of the expectation values of observable quantities should be described by a temperature-dependent pseudo-Hamiltonian $H_p$ (now commuting with reduced density matrix) which contains the information of the thermal inhomogeneity across the system and avoids the difficult problem of scattering from each temperature step. While we largely follow the framework described in detail in \cite{gm04}, we extend our pseudo free-energy $F_p$ to incorporate the chemical potential for ensembles allowing change of particle number, which can include chemical potential inhomogeneity as well:
\bea
F_p &=&\sum_n\frac{\beta_n}{\beta_0}[H_n+H_{n,n+1} - \frac{\mu_n}{\mu_0}(\mu_0N_n)],  \cr
\rho(t_0)&=&\frac{\exp(-\beta_0F_p)}{{\rm Tr}[\exp(-\beta_0F_p)]}, \;\;\;  [F_p,\rho(t_0)]=0, 
\label{1}
\eea
where $n$ labels subsystems,   $\beta_n$ and $\mu_n$ 
are the local inverse temperature and  chemical potential of the $n$th subsystem, 
$\beta_0$ and $\mu_0$ 
are arbitrary reference inverse temperature and chemical potential (for example at the midpoint). 
Here $H_n$ is the Hamiltonian of the nth subsystem, $H_{n,n+1}$ is the coupling between subsystems  and $N_n$ is the number operator.
Note that if there are no correlations between subsystems, the density matrix can be decoupled into the product of density matrices of all subsystems and there will be no reason to consider the whole sample as one system. Also, for a given fixed temperature profile in the system, a sample in NESS and  a sample in local equilibrium share the same thermodynamic properties. From now on, for simplicity, we will focus on a sample with a `quenched' temperature profile created when the sample is attached to two leads kept at different fixed temperatures and a steady state situation has been achieved with the temperature profile. We then consider the sample in isolation but with the fixed temperature profile; this quantitatively includes the temperature-dependent part of the lead-central coupling and allows us to use the framework described above to obtain the changes in the thermodynamic properties due to the inhomogeneous temperature profile. Then we add back the leads with the temperature-independent coupling, which does not change the temperature profile anymore; this allows us to evaluate the changes in transport properties due to a fixed temperature inhomogeneity. In this way, we avoid the complexities of treating the formation of the temperature profile itself, which is unknown, and the effects of coupling of the lead and the sample, simultaneously at the same level. While this may sound like an unrealizable thought experiment, the results should align well at least qualitatively with practical experimental setups if the sample thermalizes well so that the history and details of the sample entering the non-equilibrium steady state doesn't matter. Of course the method ignores any fluctuations in the temperature profile, which we leave out for future considerations.

As a simple and concrete example of a fermionic system with temperature inhomogeneity, we consider a uniform conductor, with both ends kept at temperature $T_C$ (cold) by leads at the same temperature, one end then heated by laser pulses \cite{KVY23} or electric currents \cite{li03} to temperature $T_H$ (hot), where the difference $\Delta T =T_H-T_C$ is not necessarily small, as long as the ratio $\frac{\Delta T}{T_C}$ is small. In this case there are no chemical potential inhomogeneities maintained by external sources, and as an open system coupled to two semi-infinite reservoirs (leads), there will be no extra charge accumulation on those ends due to the blockage on the boundary. As will be shown later, a position-dependent particle number density is present because of the temperature inhomogeneity, which accounts for the transport properties. However, this is distinct from an external chemical potential inhomogeneity discussed later in the context of the Widemann-Franz Law. Following \cite{gm04}, we will choose a constant temperature 
gradient within the conductor: 
\bea
T_l=T_0\left(1-\gamma\frac{l}{N}\right), \;\;\; -\frac{N}{2} < l < \frac{N}{2}.
\label{beta-l}
\eea
so that
\bea
\beta_l\approx\beta_0\left(1+\gamma\frac{l}{N}\right)
\eea
when $\gamma$ is small.
Here $l$ is the site or subsystem index, $N$ is the total number of sites along the direction of the gradient, $\beta_0$ is the mid-point or average inverse temperature of the system. This may not be accurate for a realistic system, but this captures the qualitative behavior for most of the nearly linear temperature profiles and a more elaborate depiction of the gradients could be left for future work.
From now on, we will call $\gamma$ 
the gradient parameter. Note that even with large $\Delta T$, the gradient parameter $\gamma$ can be small for a relatively large $T_0$.
To sustain a temperature 
gradient, scattering of the fermions might be required, e.g. impurity scattering and/or electron-phonon scattering. These scatterings are crucial to form the gradients and to bring fermions into steady state, but once the gradients are formed, the leading order effects are from thermal inhomogeneities (which are included in the pseudopotential). As a result, we choose non-interacting electrons (quasiparticles) on a static lattice with a fixed temperature profile as our starting point and treat the gradients as the only perturbations. The contributions from scatterings could be added back in the future and they will not undermine the effects caused by thermal inhomogeneities. 

We start with a nearest neighbor tight-binding Hamiltonian as a toy model and choose the on-site energy to be equal to the average chemical potential of the system, so that our orignal free energy only contains hopping terms and the Fermi level is at the mid-point of the conducting band. We consider effectively one-dimensional (1d) channels in a quasi-1d mesoscopic nanowire and ignore the channel mixing for simplicity. Note that for real low-dimensional system where the interactions are more important and the Fermi Liquid description breaks down, one would require an interacting model. However, as far as the effects of thermal inhomogeneity is concerned, the generalization within our framework should be straightforward. With the temperature profile given above, the temperature-independent part of the free energy with hopping strength $t$ is 
\be
H^0= -\sum_l t \;(a^{+}_{l+1}a_l+a^{+}_la_{l+1})
\label{4}
\ee
with a dispersion relation $\xi_k=-2t\cos kd$, where $d$ is the lattice spacing. The pseudo scattering term generated by the temperature gradient becomes
\bea
H^{scatt}_{\gamma}= -\frac{\gamma}{N}\sum_l t \; l(a^{+}_{l+1}a_l+a^{+}_la_{l+1}) .
\label{5}
\eea
It can be seen that the effects of the temperature gradient are depicted by mapping to a pseudo scattering process, where the hopping strength is modified to carry a pre-factor $1+\frac{\gamma l}{N}$, becoming dependent on temperature gradient and position.

\section{Green’s Functions and Dispersion Relation}
\begin{figure}
\includegraphics[angle=0,width=0.45\textwidth]{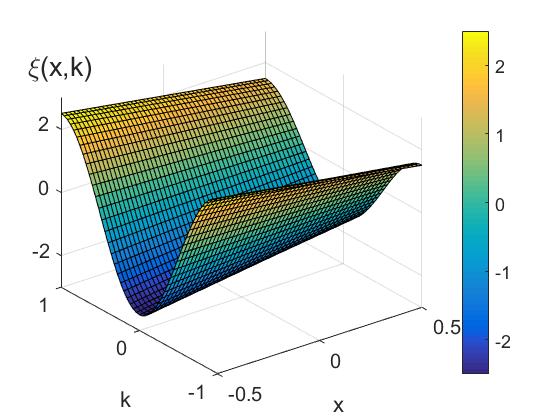}
\caption{ Momentum and position dependent dispersion due to a constant temperature gradient $\gamma=0.5$. The real space coordinate $x$ is in units of the length of the wire, the $k$-space coordinate is in units of $\frac{\pi}{d}$, where $d$ is the lattice spacing. The free energy $\xi$ is in units of the hopping parameter $t$. The $\xi=0$ free energy plane corresponds to the Fermi level.}
\label{Fig-dispersion}
\end{figure}

The time-ordered Green’s function for Wannier fields is written as
\be
G(1,2)=-iTr[\rho(t_0)\mathcal{T}a_{F_p}(1)a^{+}_{F_p}(2)]
\label{6}
\ee
where $\mathcal{T}$ is the time-ordering operator.
The first order correction to the time-ordered Green’s function coming from the pseudo scattering induced by the temperature gradient  
is given by
\bea
\delta^{(1)}G_{\gamma}(l,l',k;\tau,\tau')=i\frac{\gamma}{2N}\int d\tau_1\sum_q\delta_{q,0}\; t\;\delta \tilde{G}_{\gamma} 
\label{7}
\eea
where
\bea
&& \delta \tilde{G}_{\gamma}(l,l',k;\tau,\tau')  \cr
&&=e^{iql'd} \left[-i \left(l'e^{-ikd} + (l'-1)e^{-i(k-q)d} \right) \right. \cr
&&\left.-\left(e^{-ikd}+e^{-i(k-q)d} \right)\partial_q \right] K^0(k,k-q; \tau, \tau_1,\tau')\cr
&&- e^{iqld}  \left[i \left(le^{ikd}+(l-1)e^{-i(k+q)d} \right) \right. \cr
&&\left.+\left(e^{ikd}+e^{-i(k+q)d} \right)\partial_q \right] K^0(k+q,k; \tau, \tau_1,\tau') .
\eea
Here we defined
\bea
K^0(k_1,k_2; \tau, \tau_1,\tau') \equiv G^0(k_1,\tau-\tau_1)G^0(k_2,\tau_1-\tau').
\eea
After the integral over $q$ and conversion to frequency space, one can extract the retarded self-energy up to linear order in the gradient parameter by comparing with Dyson’s equation $G^{R}=G^{0R}\Sigma^{R}G^{R}$,
\bea
\Sigma^R_{\gamma}(l,l',k,\epsilon)=-\frac{\gamma}{N} \left(\frac{l+l'}{2}\xi_k+t \cos kd \right) .
\label{10}
\eea
Note that the retarded self-energy is purely real, which means that the quasiparticles do not decay.
The corresponding retarded Green’s function is
\bea
G^{R}_{\gamma}=\frac{1}{\epsilon-[(1-\frac{\gamma}{N}\frac{l+l'}{2})\xi_k-\frac{\gamma}{N}t \cos kd]+i\delta} .
\eea
In the large $N$ limit, the second term in the renormalized dispersion due to the temperature gradient is typically much smaller than the first term, thus we drop it here. Finally we have the renormalized dispersion to first order accuracy in the gradient parameter and in the large $N$ limit at the $l$th site,
\bea
\xi_{\gamma}(k)&=&\left(1-\frac{\gamma}{N}l \right)\xi_k .
\label{12}
\eea
Figure \ref{Fig-dispersion} shows how the dispersion depends on the momentum as well as the position coordinates in the continuum limit, due to the temperature gradient. 
Thus even the states of non-interacting particles cannot be labeled by momentum alone anymore, explicitly depending on the space coordinates as well as the parameter $\gamma$. Since $\xi_{\gamma}(k)$ appears in the fermion distribution function, both thermodynamic and transport properties can have highly non-linear $\gamma$-dependence.    
\begin{figure}
\includegraphics[angle=0,width=0.45\textwidth]{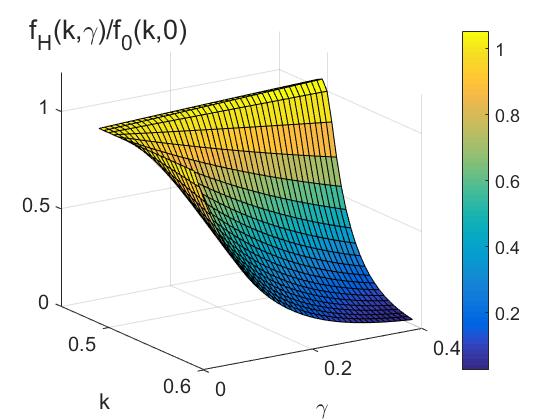}
\caption{Fermi function $f_H(k,\gamma)$ at the hot end of the sample $T_H=T_0/(1-\gamma/2)$, compared to $f_0(k,\gamma=0)$ at the midpoint temperature $T_0=300$ K, as a function of the gradient parameter $\gamma$. Momentum $k$ is in units of $\frac{\pi}{d}$, where $d$ is the lattice spacing. }
\label{Fig-Fh}
\end{figure}
Figure \ref{Fig-Fh} shows the effect of the gradient parameter $\gamma$ on the Fermi function.  

With the position and $\gamma$-dependent retarded Green’s functions, the local density of states is given by
\bea
n_{\gamma l}(\epsilon)&=&-\frac{1}{\pi}{\rm Im} G^{R}_{\gamma l}(\epsilon)=\frac{1}{N}\sum_k\delta\left(\epsilon-(1-\frac{\gamma}{N}l)\xi_k\right)\cr
&=&\frac{1}{1-\frac{\gamma}{N}l}n_0\left(\frac{\epsilon}{1-\frac{\gamma}{N}l}\right) 
\label{13}
\eea
where $ n_0$ is the midpoint density of states.
In our previous work on phonons, we showed that the consequence of temperature inhomogeneity is a temperature dependent particle number distribution, but the boson dispersion is not altered. In contrast for fermionic systems, change in the dispersion relation is significant and can be measured directly, in addition to the  local density of states. The novel fermion dispersion renormalization stems from the structure of fermionic fields and predicts nontrivial non-linearity in transport phenomena as we show below.

In the NEGF formalism of calculating current, lesser and greater Green’s functions at the lead-central interface are needed. We write them down based on the renormalized dispersion relation:
\bea
G^{<}_{\gamma} &=&i2\pi N^0\left((1-\frac{\gamma}{N}l)\xi_k\right)\delta\left(\epsilon-(1-\frac{\gamma}{N}l)\xi_k\right) \cr
G^{>}_{\gamma} &=&-i2\pi \left(1-N^0((1-\frac{\gamma}{N}l)\xi_k)\right) \cr 
&\times &\delta \left(\epsilon-(1-\frac{\gamma}{N}l)\xi_k \right) 
\label{G-LG}
\eea
where $ N^0$ is the Fermi function at the mid-point temperature $T_0$. \\

\section{Electrical and Thermal Current}
While we focused on the thermodynamic properties of systems with quenched thermal gradients so far, the attachment of two leads, each with the same temperature as the corresponding end of the nanowire, introduces trivial temperature-independent coupling to the central region and allows us to study transport properties. We employ NEGF formalism to calculate both particle and energy current \cite{LW07, Mingo06}, which mainly evaluate the particle number or energy change rate at the edge of leads due to the coupling with the central system, prepared to sustain the thermal gradients from earlier times. For simplicity and to highlight the effects of the temperature gradient, we choose the simplest model of the conductor; we assume an effectively one-channel version of a tight-binding model where the general particle and energy currents $J_N$ and $J_E$ can be respectively written as
\bea
J_N&=&\int\frac{d\epsilon}{4\pi} [\int\frac{dk_L}{2\pi}\frac{dk_C}{2\pi} (G^{(C)>}_L(k_C,\epsilon)\Sigma^<_L(k_L,\epsilon) \cr
&-&G^{(C)<}_L(k_C,\epsilon)\Sigma^>_L(k_L,\epsilon))]-[L\to R] ,\cr
J_E&=&\int\frac{d\epsilon}{4\pi} [\int\frac{dk_L}{2\pi}\frac{dk_C}{2\pi}\frac{\xi_{k_L}}{2}(G^{(C)>}_L(k_C,\epsilon)\Sigma^<_L(k_L,\epsilon) \cr
&-&G^{(C)<}_L(k_C,\epsilon)\Sigma^>_L(k_L,\epsilon))]-[L\to R].
\label{15}
\eea
Here the superscript (C) on the Green's function $G$ refers to the central region (the wire), with subscripts L and R referring to the ends connecting the left and right leads, respectively. The subscripts L and R on the self energy $\Sigma$ refer to the left and right leads, respectively. Note that in the general formalism, to derive the expressions of the currents above, the coupling between leads and the central system is crucial. Therefore Green’s functions used are coupled (non-equilibrium) Green’s functions, which contain both the many body effects inside the central system and the coupling between the leads and central region. To achieve more accurate results, self-consistent procedures were applied in some works to treat the many body effects and the lead-central coupling simultaneously \cite{LW07, Mingo06}. However, to illustrate the effects of the thermal inhomogeneity, we assume that in a long enough wire in the absence of any strong many-body effects in the central region or in the leads, the global thermal inhomogeneities across the system are more important than the lead-central coupling that only exists at the lead-central interface. Since in our toy model  the interface is essentially only one slice thick, we make the approximation of neglecting the lead-central coupling and the coupled (non-equilibrium) Green’s functions reduce to equilibrium ones. Since the lead-central coupling here doesn't play any role in the formation of thermal gradient, we assume it is independent of the gradient parameter. In real systems, our assumptions should remain approximately valid as long as the interface is sufficiently thin and thermalizes quickly. Thus we expect that our results will remain qualitatively valid (as far as the effects of thermal inhomogeneity is concerned) compared to any future results from self-consistent coupled (non-equilibrium) Green’s functions. 

The lead’s self-energy is proportional to the lead’s non-interacting Green’s function with a proportionality factor related to the coupling between the lead and the central region. As a result, the evaluations of currents still heavily depend on the properties of the lead-central coupling even though we simplified the central Green's function to exclude that part. However, as we will show, if one pursues the ratios of the transport coefficients between nonlinear response regime and linear response regime, this common factor will be canceled; the corrections to the linear response regime don't depend on the explicit form of lead-central coupling.

Using the relevant Green’s functions in the central region obtained above and after the $k$-integral we obtain particle current due to temperature gradient related to the  Seebeck coefficient and energy current due to temperature gradient related to the thermal conductivity: 
\bea
&&J^{\gamma}_N \propto\int d\epsilon\; n_0(\epsilon) \left[ \frac{1}{1+\frac{\gamma}{2}} n_0 \left (\frac{\epsilon}{1+\frac{\gamma}{2}}\right) \delta N_1^0 \right. \cr
&&\;\;\;\;\;+ \left.n_0 \left((1-\frac{\gamma}{2})\epsilon \right) \delta N_2^0 \right], \cr
&&J^{\gamma}_E\propto \int d\epsilon \; n_0(\epsilon) \left[ \frac{\epsilon}{1+\frac{\gamma}{2}} n_0 \left(\frac{\epsilon}{1+\frac{\gamma}{2}} \right)\delta N_1^0 \right.\cr
&&\;\;\;\;\; + \left.\left((1-\frac{\gamma}{2})\epsilon \right) n_0 \left((1-\frac{\gamma}{2})\epsilon \right) \delta N_2^0 \right]
\label{16}
\eea
where 
\bea
&&\delta N_1^0=N^0 \left((1-\frac{\gamma}{2})\epsilon \right) -N^0(\epsilon) \cr
&& \delta N_2^0 = N^0 \left((1-\frac{\gamma}{2})\epsilon \right)-N^0 \left((1-\frac{\gamma^2}{4})\epsilon \right). 
\eea

\section{Transport Coefficients and Lorenz Number}
\begin{figure}
\includegraphics[angle=0,width=0.45\textwidth]{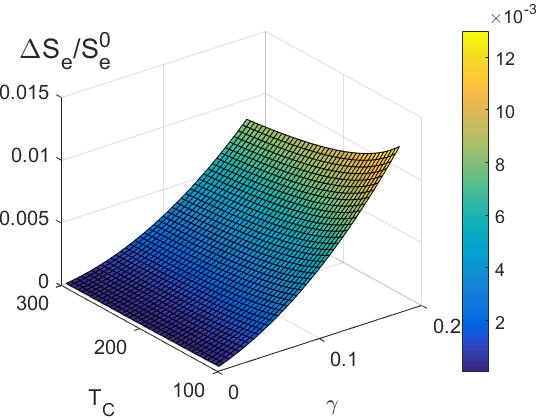} 
\caption{The non-linear increase in the ratio $\frac{\Delta S_e}{S^0_e}$ with the gradient parameter $\gamma$, showing the deviation of thermopower from linear response for finite $\gamma$. $T_C$ is the temperature of the colder end of the device.  }
\label{Fig-8S_S0_integ}
\end{figure}

We start from the three-dimensional density of states; setting the momentum along the direction perpendicular to the propagation to be zero gives us the form of density of states for an effective one-dimensional channel. Using a parabolic dispersion and in the continuum limit, the density of states becomes
\be
n_0(\epsilon)\propto \frac{1}{\sqrt{\epsilon+2t}}.
\label{18}
\ee
Note that although the transport coefficients themselves explicitly depend on the density of states, the qualitative behavior of the ratios beween transport coefficients, e.g. corrections to linear response regime, do not have any significant dependence on the form of the density of sates.
We propose to consider the non-linear response versions of various transport coefficients as follows,
\bea
S^{\gamma}_{e,nl}\equiv\frac{J^{\gamma}_N}{\Delta T}, \;\;\;
\kappa^{\gamma}_{nl}\equiv\frac{J^{\gamma}_E}{\Delta T}, 
\eea
where $S^{\gamma}_{e,nl}$ is the thermopower and  $\kappa^{\gamma}_{nl}$ is the thermal conductance. Note that the transport coefficients defined above are “conductance”, not “conductivity”, but the length dependence cancels out when evaluating the ratio $S^{\gamma}_{e,nl}/S^0_{e,nl}$ or $\kappa^{\gamma}_{nl}/\kappa^0_{nl}$,  where the superscript 0 corresponds to the limit $\gamma\to 0$. We then define and calculate the following ratio to reflect the
corrections to the linear response thermopower due to the finite temperature gradient:
\bea
\frac{\Delta S_e}{S^0_e}&=&\frac{S^{\gamma}_{e,nl}-S^0_{e,nl}}{S^0_{e,nl}},
\label{20}
\eea
The result shown in Figure \ref{Fig-8S_S0_integ} reveals the non-linear increasing behavior of the thermopower with the gradient parameter $\gamma$.

The Widemann-Franz Law for Fermi-liquids says that the ratio $\kappa/T_0\sigma$ is a constant, equal to $\pi^2/3$. 
\begin{figure}
\includegraphics[angle=0,width=0.45\textwidth]{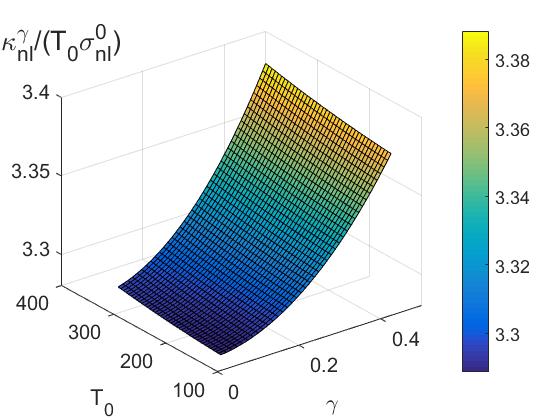} 
\caption{The increase in the ratio  $\kappa^{\gamma}_{nl}/(T_0\sigma^0_{nl})$ with the gradient parameter $\gamma$, showing the generalization of the Widemann-Franz Law for $\gamma\ne 0$. $T_0$ is the midpoint temperature.  }
\label{Fig-WFL}
\end{figure}
To check if the linear response limit is properly taken into account in this formulation, we need to consider the electrical conductance within the same framework. We therefore assume a fictitious chemical potential gradient $\eta$ given by $\mu_l=\mu_0(1-\eta l/N)$ in the absence of temperature inhomogeneity similar to the temperature gradient given in (\ref{beta-l}), and repeat the perturbation calculation within the same framework, then find the electrical conductance from the non-equilibrium particle current. In this case
in the $\gamma\to 0, \; \eta\to 0$ limit,
\bea
\frac{\kappa^0_{nl}}{T_0\sigma^0_{nl}}&=&\frac{\mu_0}{T^2_0}\frac{\partial_{\gamma}J^{\gamma}_E|_{\gamma=0}}{\partial_{\eta}J^{\eta}_N|_{\eta=0}} \cr
&=&\frac{\mu_0}{T^2_0}\frac{\int d\epsilon\; n_0^2(\epsilon)\epsilon^2\frac{dN^0(\epsilon)}{d\epsilon}}{\mu_0\int d\epsilon \; n^2_0(\epsilon)
\frac{dN^0(\epsilon)}{d\epsilon}}\cr
&=&\frac{\mu_0}{T^2_0}\frac{n^2_0(\epsilon_F)}{n^2_0(\epsilon_F)}\frac{T^2_0\frac{\pi^2}{3}}{\mu_0} = \frac{\pi^2}{3}
\eea
and the Lorenz number is restored to the constant $\pi^2/3$ in the linear response regime. Thus, our formulation generalizes the Widemann-Franz law, which is valid for infinitesimal temperature difference across the conductor, to finite temperature gradients across the system as shown in Figure \ref{Fig-WFL}.

\section{Summary and Discussion}
We consider a simple nanowire with a linear thermal gradient characterized by the parameter $\gamma$, with charge and thermal currents generated by the temperature difference between the two ends of the wire ignoring any contribution to the thermal conductance from phonons, which could be achieved by introducing surface disorder \cite{li03, boukai08, hochbaum08, chen08, lim12, heron09, blanc13, blanc14, ma17, mm18, mm19, am19}. The two ends have fixed temperatures $T_C=T_0(1-\gamma/2)$ and $T_H=T_0(1+\gamma/2)$ where $T_0$ is the temperature of the midpoint. The temperature inhomogeneity makes application of non-equilibrium field theory difficult even for such a simple system because Wick's theorem cannot be applied directly. We adapt a framework developed earlier for general systems with thermal inhomogeneity. We show that in contrast to a bosonic system considered earlier,  the temperature inhomogeneity in a fermionic system gives rise to a position and $\gamma$ dependent dispersion relation, and therefore a Fermi distribution that explicitly depends on the temperature gradient and space coordinates, affecting thermodynamic as well as transport properties of the system in highly non-trivial ways. In particular, we evaluate the non-linear number current $J^{\gamma}_N$ as well as the thermal current $J^{\gamma}_E$ as functions of the gradient parameter $\gamma$ and evaluate the non-linear thermopower $S^{\gamma}_{e,nl}$ (the Seebeck coefficient) and the thermal conductance $\kappa^{\gamma}_{nl}$. In the same framework, one can consider a chemical potential inhomogeneity. This allowed us to show how the Widemann-Franz Law can be generalized for finite $\gamma$ and is recovered exactly in the linear response regime.

In this work, while we study the simplest model that illustrates the effects of thermal inhomogeneity on thermodynamic as well as transport properties, the framework used here should be more generally valid beyond effectively one-dimensional non-interacting systems or the specific temperature profile considered here. In particular, it would be interesting to see if the formulation can be adapted to many-body systems with thermal inhomogeneities, including many-body localization \cite{NH15}. Additionally, in 2D systems where time reversal symmetry is broken, our framework might become a useful tool to explore thermal hall effect \cite{KNL10} beyond Kubo formula.

\section{CRediT authorship contribution statement}
\textbf{Yuan Gao:} Conceptualization, Formal analysis, Investigation, Writing. \textbf{K.A. Muttalib:} Conceptualization, Supervision, Writing.

\section{Declaration of Competing Interest}
The authors declare that they have no known competing financial interests or personal relationships that could have appeared to influence the work reported in this paper.
 

\end{document}